\documentclass[conference]{IEEEtran}
\IEEEoverridecommandlockouts

\usepackage{cite}
\usepackage{amsmath,amssymb,amsfonts}
\usepackage{algorithmic}
\usepackage{graphicx}
\usepackage{textcomp}
\usepackage{xcolor}
\usepackage{float}
\usepackage{stfloats}
\usepackage[hidelinks]{hyperref}
\usepackage{comment}

\def\BibTeX{{\rm B\kern-.05em{\sc i\kern-.025em b}\kern-.08em
    T\kern-.1667em\lower.7ex\hbox{E}\kern-.125emX}}
\begin{document}

\title{
A 21–24 GHz Low-Phase-Noise mmWave VCO with Third-Harmonic Expansion using a Triple-Coupled Transformer based Tank
}

\author{\IEEEauthorblockN{Narahari N. Moudhgalya}
\IEEEauthorblockA{
\textit{IIIT Hyderabad}\\
Hyderabad, India \\
narahari.moudhgalya@research.iiit.ac.in}
}



\maketitle

\begin{abstract}
This work presents the design and analysis of a 
sixth-order triple-coupled 
transformer-based tank,
enabling
third-harmonic expansion for mmWave VCOs.
Unlike conventional fourth-order tanks,
the proposed tank inherently supports three resonance modes, 
enabling 
wideband third-harmonic expansion 
without additional low-Q 
switched-capacitor
tuning elements.
In contrast to
conventional class-F23 designs, 
the proposed VCO removes the head resonator and adopts a noise circulating core to 
maintain 
low phase noise with reduced area.
Implemented in TSMC 65-nm CMOS,
post-layout
simulation  results demonstrate a 21.03–23.99 GHz (13.5\%) tuning range, 
minimum
phase noise of –116.25 dBc/Hz 
at 1 MHz
offset, 
and peak FoM/FoMT/FoMA of 195.86/198.24/212.31 dBc/Hz
while consuming
5.4 mW and occupying 0.02268 mm2.
\end{abstract}


\section{Introduction}

There is growing interest in highly integrated radar transceivers operating at millimeter-wave (mmWave) frequencies. 
In such systems, low phase-noise (PN) voltage-controlled oscillators (VCOs) are essential, as spectral purity directly affects modulation accuracy and overall system robustness \cite{two_port_2018, two_port_2019, two_port_2020_TCAS1_RuiMartin_current_reuse}. 
In CMOS 
VCOs,
PN is primarily governed by resonator losses and by the up-conversion of device noise within the active core. 
Improving resonator impedance while limiting noise conversion therefore remains central to low-PN VCO design.
Two broad strategies have been widely explored.
One approach increases the effective impedance of the tank using multi-port or multi-resonant structures
\cite{two_port_2006, two_port_2018, two_port_2019, two_port_2020_TCAS1_RuiMartin_current_reuse, two_port_2020_RFIC, two_port_2022_RuiMartin_mmWave_RLCM}. 
Another approach shapes the oscillation waveform by introducing 
impedance peaks at harmonic frequencies
to modify
the impulse sensitivity function (ISF) \cite{Hajimiri_658619}
and reduce noise up-conversion 
\cite{Hegazi, hr_classF, Murphy, hr_classF-1, hr_1byf_upconversion, hr_classF23_explicit_CM, hr_classF23_optimal_Q, hr_inverse_classF23, hr_VCO_Tutorial, hr_trifilar, hr_classF-1_2023}.

In many 
harmonic-rich (HR) VCO implementations,
fourth-order resonator tanks
are combined with additional tuning elements to realize
differential-mode (DM) third-harmonic 
tuning
\cite{lowQ_10130821, lowQ_9365761, lowQ_9794818}.
While 
they
enable third-harmonic control, they introduce extra tuning elements that either reduce the effectiveness of harmonic shaping or lower the overall tank quality factor (Q), 
particularly at mmWave frequencies.
An alternative approach 
is to 
employ transformer (XFMR) coupled multi-LC networks to achieve DM third-harmonic expansion.
However, this comes at the cost of increased inductor area \cite{6th_order_tank_OG_10068119}.
\begin{figure}
    \centering
    \includegraphics[width=\linewidth]{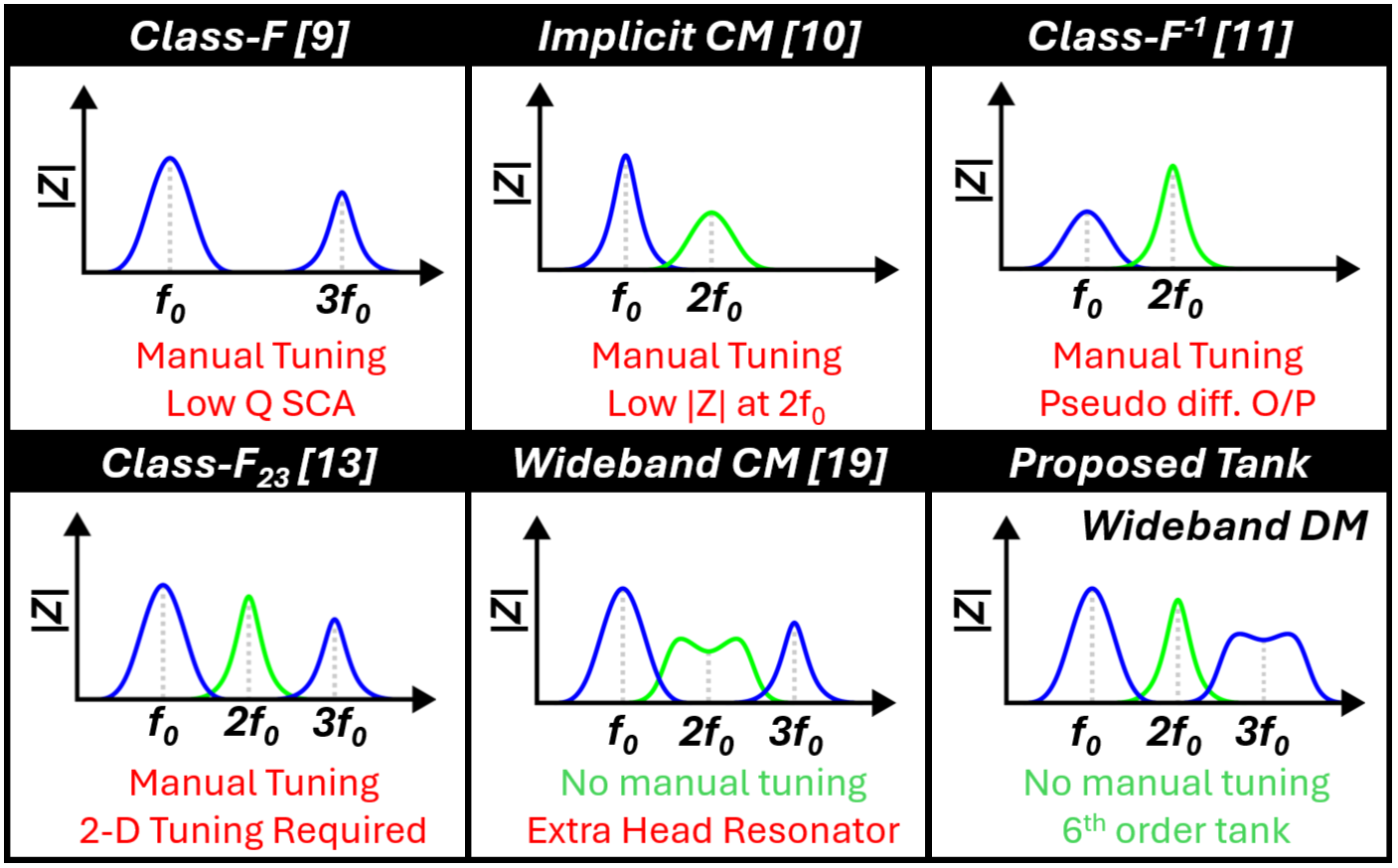}
    \caption{Comparison of recent harmonic-rich-shaping VCO Techniques}
    \label{fig:Comparison_tank}
\end{figure}
To address this issue, a sixth-order tank can be realized using a triple-coupled XFMR,
which inherently exhibits three resonance modes.
By aligning the higher-order modes around the third harmonic, 
wideband third-harmonic expansion can be achieved without introducing auxiliary low-Q switched-capacitor arrays (SCA).
However, a systematic closed-form analysis of sixth-order XFMR-based resonator tanks has remained limited in the literature. The sixth-order topology reported in \cite{classF23_6th_order} incorporates an additional tail-resonator to broaden the common-mode (CM) response, which increases circuit complexity and area. Moreover, certain sign inconsistencies observed in the frequency expressions of \cite{classF23_6th_order} yield non-physical (NaN) solutions for two of the three reported resonance frequencies.
In this work, a generic closed-form analysis of the sixth-order triple-coupled XFMR tank is developed, 
resulting in consistent expressions for all three DM resonance frequencies.
As shown in Fig.~\ref{fig:Comparison_tank}, the higher-order resonance modes are then positioned around 
the third harmonic,
resulting in wideband third-harmonic expansion without auxiliary low-Q tuning elements.
To avoid the area overhead associated with second-harmonic tuning structures, 
the proposed VCO removes the additional resonator 
and adopts a noise circulating (NC) core \cite{NC_8637955}. By reducing the portion of device noise that reaches 
the tank, the NC topology enables comparable PN 
performance while improving area efficiency.
%

The remainder of the paper is organized as follows. 
Section~II reviews relevant 
prior work. 
Section~III presents a generic analysis of the proposed sixth-order 
XFMR-based
tank and discusses third-harmonic
expansion. 
Section~IV describes a 21-24~GHz mm-wave VCO implementation and 
post-layout simulations
results. 
Finally, Section~V concludes the paper.

\section{Review of Harmonic Rich Shaping Techniques}
Fig. \ref{fig:Comparison_tank} depicts various HR shaping techniques 
for PN reduction.
These methods differ mainly 
in how the second and third harmonic resonances are introduced and how their alignment with the fundamental frequency is maintained.
~\cite{Hegazi} employs a tail resonator to prevent Q degradation of the LC tank 
when the cross-coupled transistors operate in the triode region.
This helps suppress flicker noise up-conversion but requires an additional inductor and manual alignment of the tail-resonator frequency, which increases area.
To avoid the extra inductor,
~\cite{Murphy} demonstrates
implicit CM resonance within a single 
resonator tank. 
Although this improves area efficiency, the achievable CM impedance and Q are inherently limited 
because CM currents experience magnetic flux cancellation.

As shown in Fig. \ref{fig:Comparison_tank}, XFMR-based tanks have been widely used to introduce harmonic impedance peaks for PN improvement. In \cite{hr_classF}, a class-F oscillator adds a third-harmonic resonance to shape the ISF. In contrast, the class-F$^{-1}$ topology in \cite{hr_classF-1} realizes a higher impedance at the second harmonic than at the fundamental, with simultaneous shaping of the fundamental and second-harmonic components. The class-F$_{23}$ architecture in \cite{hr_classF23_explicit_CM} further incorporates both second and third-harmonic resonances, typically through an SCA that tunes 
both CM and DM harmonics 
%
Despite their effectiveness, these designs depend on explicit harmonic tuning to maintain the intended relationship between the fundamental and harmonic frequencies over the tuning range (TR). This usually requires additional SCAs or related tuning elements. At mmWave frequencies, the added parasitics reduce the effective tank Q and limit the achievable PN performance.

To avoid explicit harmonic tuning, several architectures have been proposed that 
expand
the CM second-harmonic resonance, including the use of a head resonator \cite{head_resonator_9365761}, 
dual CM resonators \cite{Dual_CM_9794818}, and 
transformer-based resonators incorporating auxiliary coils \cite{aux_coil_10130821}. 
However, these solutions often weaken the harmonic shaping mechanism by 
intentionally lowering the impedance or Q of the DM third-harmonic resonance \cite{head_resonator_9365761}, 
relying on switched transformers \cite{switched_XFMR_10274730}, 
or introducing additional SCAs \cite{aux_coil_10130821}. 
Multi-LC tank architectures with transformer coupling are reported in \cite{lowQ_10130821, classF23_6th_order ,multi_LC_10068119} to expand both CM and DM 
resonances.
However, 
the use of multiple inductors leads to a substantial increase in chip area.
This highlights the need for achieving third-harmonic expansion without additional tuning elements, while keeping the area low, to obtain Class-F operation across the TR.

In the proposed approach, the sixth-order triple-coupled XFMR tank is designed to realize third-harmonic expansion 
without employing dedicated second-harmonic tuning structures, which reduces area. 
To compensate for the removal of the second-harmonic resonator, a NC core is employed, which reduces the device noise coupled into the tank,
hence maintaining the PN performance.
The following section describes the tank structure and its analysis.

\section{Analysis of the proposed XFMR based tank}
\begin{figure}
    \centering
    \includegraphics[width=1\linewidth]{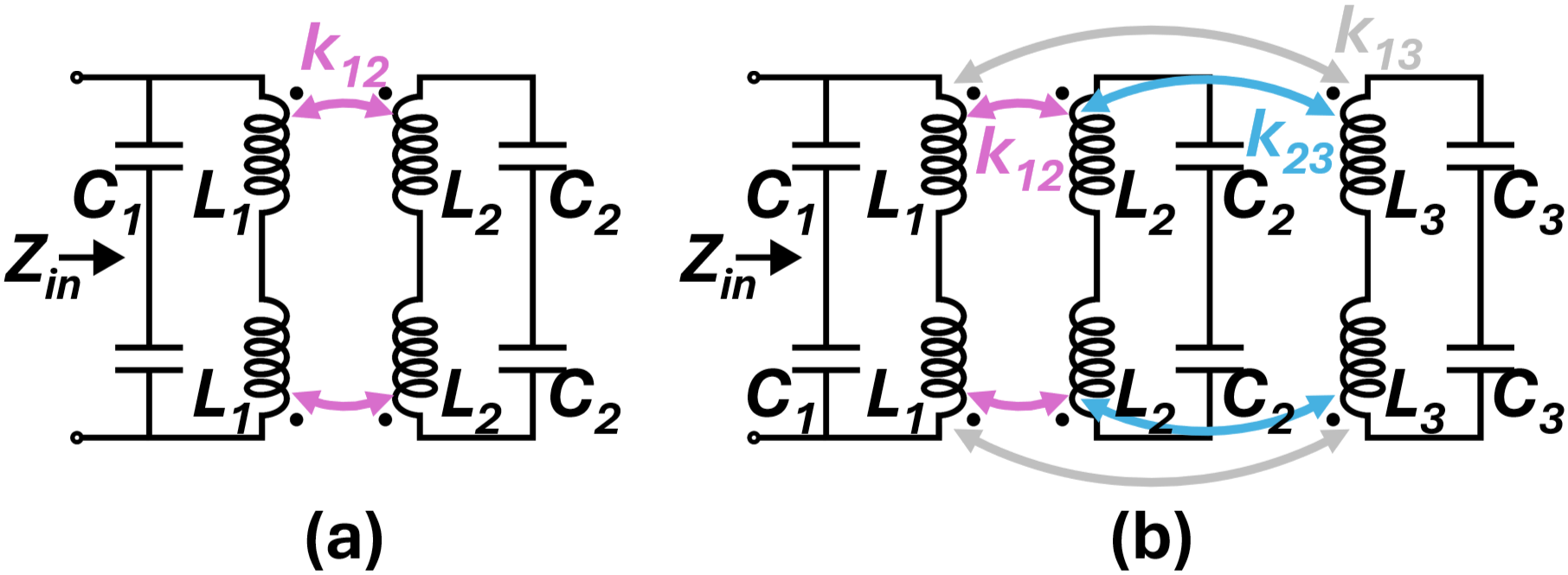}
    \caption{(a) Conventional fourth-order XFMR tank and (b) proposed sixth-order triple-coupled XFMR tank}
    \label{fig:ProposedTank}
\end{figure}
\begin{figure}
    \centering
    \includegraphics[width=\linewidth]{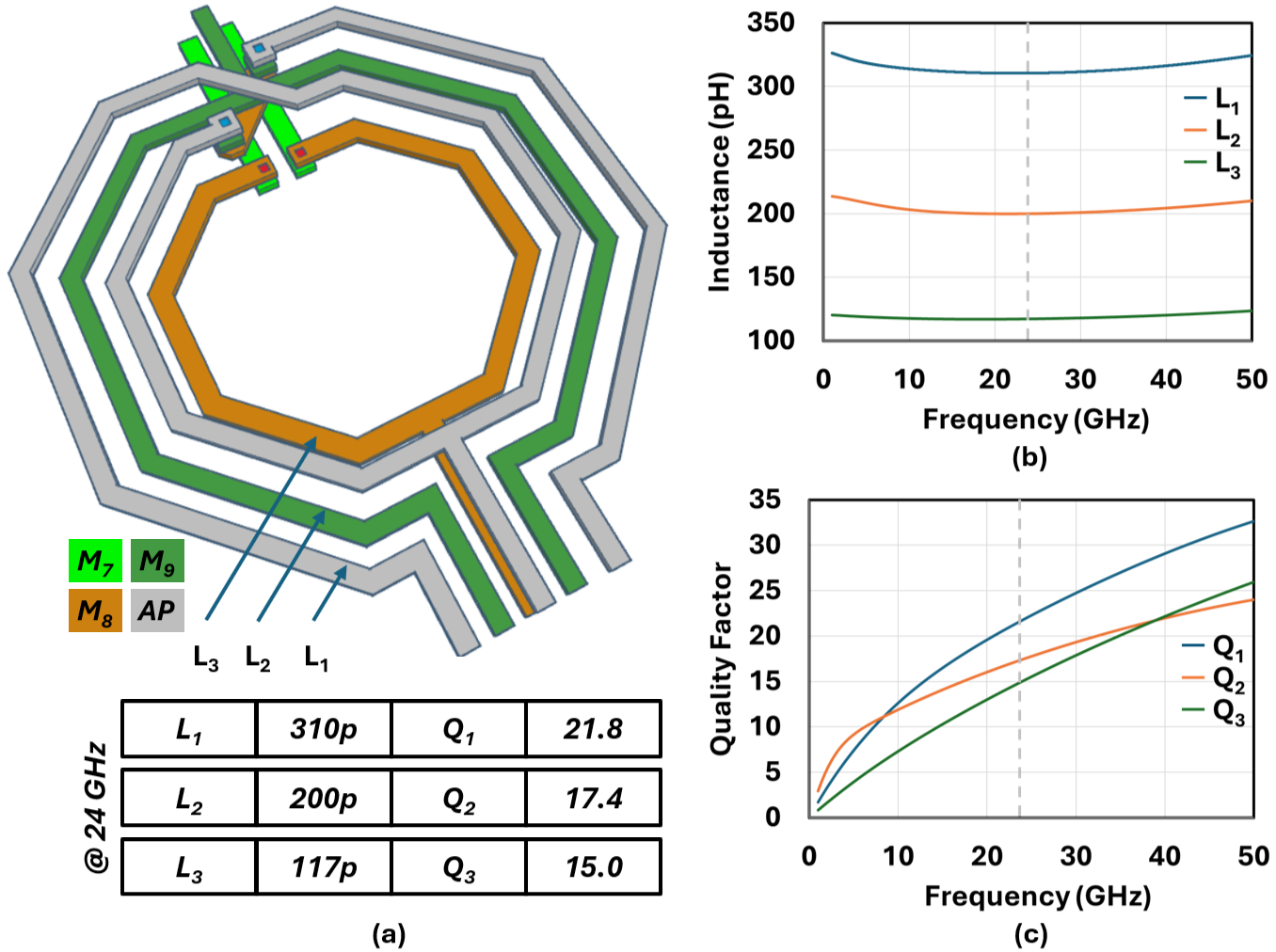}
    \caption{(a) Transformer layout and Cadence EMX simulation results - (b) Inductance and (c) Quality Factor}
    \label{fig:XFMR}
\end{figure}
As illustrated in Fig.~\ref{fig:ProposedTank} and \ref{fig:XFMR}, 
the proposed topology employs a triple-coupled XFMR-based LC network. 
In conventional fourth-order tanks, harmonic alignment is typically achieved using additional low-Q SCAs, which introduce loss and reduce the effective tank Q. In the proposed structure, the sixth-order response arises directly from magnetic coupling 
among the three inductive branches, and no separate harmonic-tuning SCAs are used.
The resonance characteristics of this coupled network are analyzed in the following subsections.

\subsection{DM Impedance of the triple-coupled XFMR tank}

Under DM excitation, the coupled tank can be represented using its half-circuit model.
In this model, the series impedance of each inductor branch is given by $Z_i = sL_i + r_i$, where $L_i$ and $r_i$ denote the self-inductance and equivalent series resistance of the $i$-th branch, respectively. 
Capacitors $C_1$, $C_2$ and $C_3$ denote the effective $C_{DM}$ in each branch. 
The magnetic coupling between the coils is captured through the mutual inductances $M_{ij}$.
Applying KVL to the inductor branches
leads to \eqref{eq:V_I_matrix} and \eqref{eq:load_relation}.
For simplicity of the analysis, the tank is assumed to be lossless ($r_i$ = 0).
Solving \eqref{eq:V_I_matrix} and \eqref{eq:load_relation} for the input impedance $Z_{\text{in}}$ results in the expression given in \eqref{eq:Zin}, where the effective impedance $Z_{\text{eff}}$ is defined in \eqref{eq:Zeff}.
Equation \eqref{eq:Zin}
is the generic $Z_{in}$ equation for a triple-coupled XFMR tank, and it
reveals that the resulting tank exhibits a sixth-order response, in contrast to the fourth-order behavior typically observed in conventional class-F oscillators \cite{hr_classF}.
\begin{figure}[!t]
    \centering
    \includegraphics[width=\linewidth]{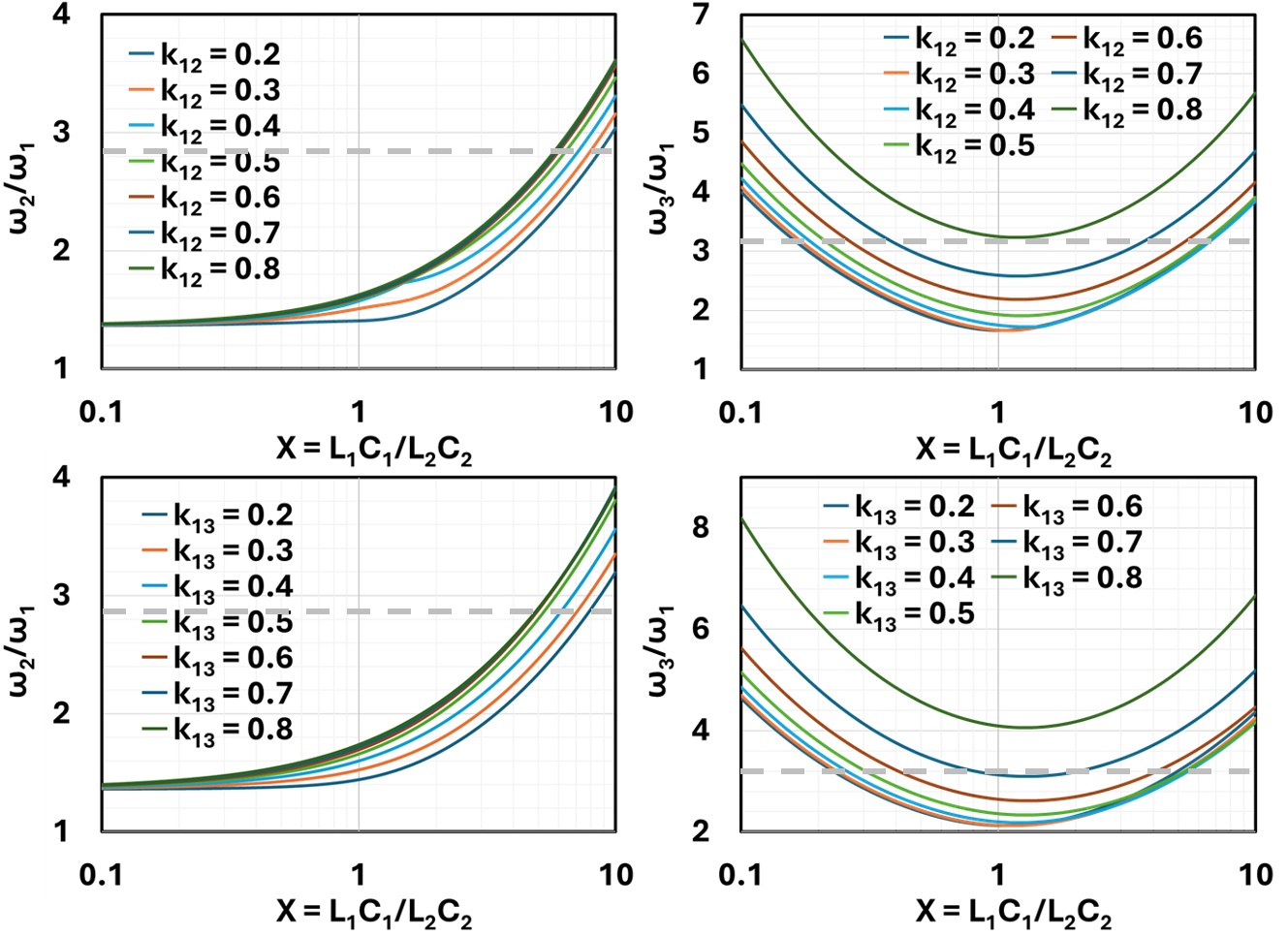}
    \caption{Variation of frequency ratios with \textit{(a)-(b)} $k_{12}$ and \textit{(c)-(d)} $k_{13}$}
    \label{fig:freq_ratio_variation}
\end{figure}
\begin{align}
\begin{pmatrix}
V_1 \\
V_2 \\
V_3
\end{pmatrix}
&=
\begin{pmatrix}
j\omega Z_1 & j\omega M_{12}   & j\omega M_{13} \\
j\omega M_{12}   & j\omega Z_2 & j\omega M_{23} \\
j\omega M_{13}   & j\omega M_{23} & j\omega Z_3
\end{pmatrix}
\begin{pmatrix}
I_1 \\
I_2 \\
I_3
\end{pmatrix}
\label{eq:V_I_matrix}
\\
\begin{pmatrix}
V_2 \\
V_3
\end{pmatrix}
&=
\begin{pmatrix}
- 1/sC_2 & 0 \\
0 & - 1/sC_3
\end{pmatrix}
\begin{pmatrix}
I_2 \\
I_3
\end{pmatrix}
\label{eq:load_relation}
\end{align}
\vspace{-10pt}
\begin{align}
Z_{in}(s) = \frac{Z_{eff}(s)}{1 + s C_1 Z_{eff}(s)} \label{eq:Zin} 
\end{align}
\begin{figure*}[!b]
\hrulefill
\begin{equation}
Z_{eff}(s)
=
Z_{L1}
-
s^3
\frac{
M_{12}^2 C_2 + M_{13}^2 C_3
+ s^2 C_2 C_3
\bigl(
L_3 M_{12}^2
+ L_2 M_{13}^2
- 2 M_{23} M_{12} M_{13}
\bigr)
}{
1
+ s^2 (L_2 C_2 + L_3 C_3)
+ s^4 C_2 C_3 (L_2 L_3 - M_{23}^2)
}
\label{eq:Zeff}
\end{equation}
\end{figure*}
These results are used to extract the key characteristics of the tank in the following subsections.

\subsection{Oscillation Frequency}
\begin{figure}
    \centering
    \includegraphics[width=1\linewidth]{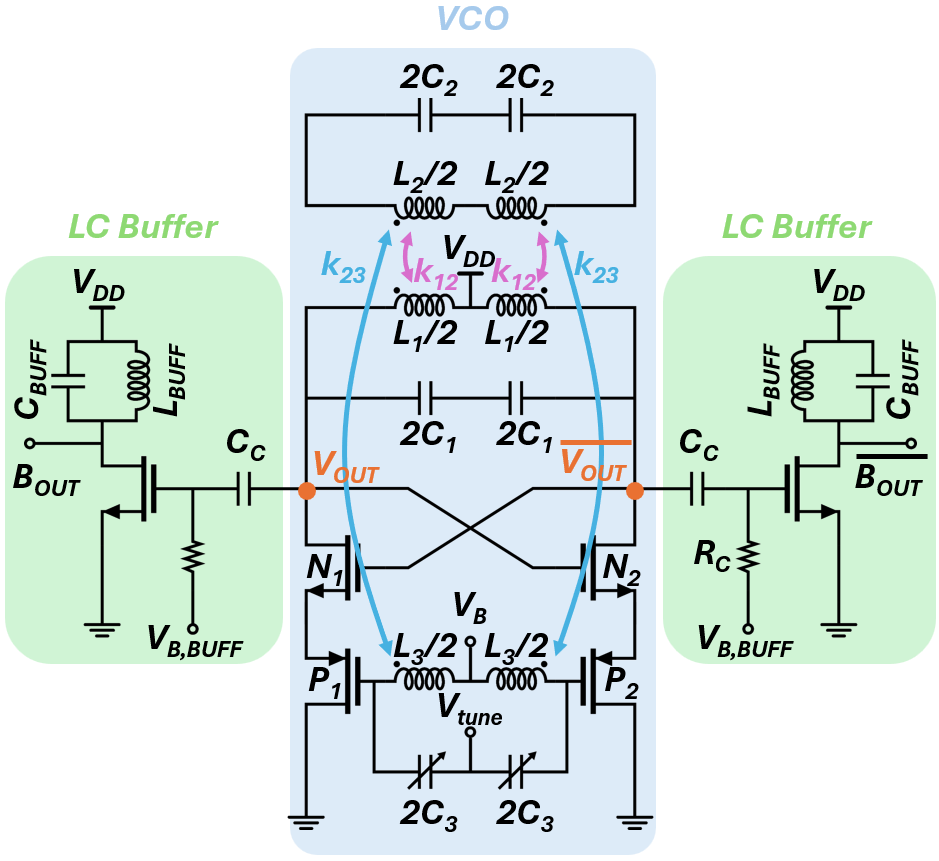}
    
    \caption{Proposed Topology with Noise Circulation Technique and sixth-order triple-coupled XFMR based tank}
    \label{fig:Proposed_Topology}
\end{figure}


The proposed triple-coupled XFMR tank supports three differential-mode (DM) oscillation
modes arising from magnetic coupling among the LC tanks.
Under lossless and high-$Q$ assumptions, the DM oscillation frequencies are obtained from the
poles of the input impedance in \eqref{eq:Zin}
This leads to a cubic characteristic equation in
$x=\omega^2$, given by \eqref{eq:characetristic_equation},
where the coefficients are expressed in terms of the uncoupled tank frequencies
$\omega_i = 1/\sqrt{L_iC_i}$ and normalized coupling coefficients $k_{ij}$, given by \eqref{eq:char_coeffs}.
\begin{equation} \label{eq:characetristic_equation}
C_3 x^3 - C_2 x^2 + C_1 x - C_0 = 0 ,
\end{equation}
\vspace{-15pt}
\begin{align} \label{eq:char_coeffs}
C_3 &= 1-(k_{12}^2+k_{13}^2+k_{23}^2)+2k_{12}k_{13}k_{23}, \notag\\
C_2 &= (1-k_{12}^2)\omega_3^2+(1-k_{13}^2)\omega_2^2+(1-k_{23}^2)\omega_1^2, \notag\\
C_1 &= \omega_1^2\omega_2^2+\omega_1^2\omega_3^2+\omega_2^2\omega_3^2, \notag\\
C_0 &= \omega_1^2\omega_2^2\omega_3^2.
\end{align}
This cubic
admits three real, positive roots
$\omega_1 < \omega_2 < \omega_3$, corresponding to one fundamental DM mode and two higher-order modes
introduced by magnetic coupling. For physically realizable three-tank LC systems with moderate coupling,
all three modes remain real and distinct. Closed-form expressions for the roots are obtained using the
standard trigonometric solution of the cubic (Cardano--Vi\`ete form), with coefficients
given in \eqref{eq:char_coeffs},
which are determined by
the circuit parameters. These roots directly correspond to the three 
DM
oscillation
frequencies of the coupled-tank system, and are given by 
\eqref{eq:wL2}, \eqref{eq:wH12} and \eqref{eq:wH22} respectively,
where $\theta$ is given by \eqref{eq:theta}.

\begin{align}
\omega_1^2 &=
\frac{
C_2
- 2 \sqrt{-3C_1C_3 + C_2^2}
\cos\!\left(\frac{\theta - \pi}{3}\right)
}{
3C_1
} \label{eq:wL2} \\[6pt]
\omega_2^2 &=
\frac{
C_2
- 2 \sqrt{-3C_1C_3 + C_2^2}
\cos\!\left(\frac{\theta + \pi}{3}\right)
}{
3C_1
} \label{eq:wH12} \\[6pt]
\omega_3^2 &=
\frac{
C_2
+ 2 \sqrt{-3C_1C_3 + C_2^2}
\cos\!\left(\frac{\theta}{3}\right)
}{
3C_1
} \label{eq:wH22}
\end{align}
\begin{align}
\theta =
\arccos&\!\left(
\frac{
27C_1^2 \omega_1^2 \omega_2^2 \omega_3^2
- 9C_1C_2C_3
+ 2C_2^3
}{
2\sqrt{\left(-3C_1C_3 + C_2^2\right)^3}
}
\right)
\label{eq:theta}
\end{align}

A closely related harmonic-frequency analysis is presented in \cite{classF23_6th_order}. However, the trigonometric solution reported there contains sign inconsistencies in the square-root and inverse-cosine terms of the cubic expression. Under certain coupling conditions, these sign differences lead to non-real modal frequencies. In this work, the cubic solution is re-derived from the characteristic equation with consistent sign conventions, resulting in physically valid DM resonance frequencies.
These frequency expressions form the basis for aligning the higher-order modes to achieve controlled third-harmonic expansion, as discussed in the next subsection.

\subsection{
Derivation of the 
passive components values
and third-harmonic expansion
}

To achieve DM 
$3^{rd}$ 
third-harmonic expansion
the two high impedance peaks, 
$\omega_2$ and $\omega_3$, 
must be 
aligned around $3\omega_1$.
By setting $\omega_2$=$\omega_3$ (i.e, $L_2C_2\text{=}L_3C_3$), 
the ratios $\omega_2/\omega_1$ and $\omega_3/\omega_1$ can be calculated versus $X\text{=}L_1C_1/L_2C_2$ for different $k_{12}$ and $k_{13}$.
It is observed that 
the ratio moves to higher values for larger $k_{23}$ and finally the higher order resonance modes disappear for the perfect coupling factor (=1). Hence, $k_{23}$ is kept relatively low ($\approx0.2-0.3$)
to preserve distinct higher-order resonances
.
Using~\eqref{eq:Zin}, the input impedance $Z_{\mathrm{in}}$ is plotted for various combinations of $k_{12}$ and $k_{13}$ using Python/MATLAB. 
Fig.~\ref{fig:freq_ratio_variation} shows the resulting trends, which
are used to determine the coupling values that provide the desired bandwidth around the third-harmonic. A bandwidth of approximately 30\% is targeted to accommodate the intended third-harmonic across TR as well as variations due to process, voltage, and temperature (PVT). Based on this analysis, the triple-coupled XFMR is designed as shown in Fig.~\ref{fig:XFMR}.
The following section discusses 
the implementation 
and post layout simulation results of the proposed topology.




\section{Implementation Details and Post Layout Simulations Results}

Fig. \ref{fig:Proposed_Topology} shows the proposed VCO topology with a triple-coupled XFMR tank and NC active core,
implemented
in TSMC 1P9M\_6X1Z1U 65-nm CMOS technology.
%
Fig. \ref{fig:XFMR} shows the XFMR layout and the Cadence EMX simulation results.
At 24~GHz,
the coils have values $L_1\text{=}300pH$, $L_2\text{=}210pH$ and $L_3\text{=}117pH$ with Q values 21.8, 17.4 and 15 respectively.
\begin{figure}
    \centering
    \includegraphics[angle=0, width=0.9\linewidth]{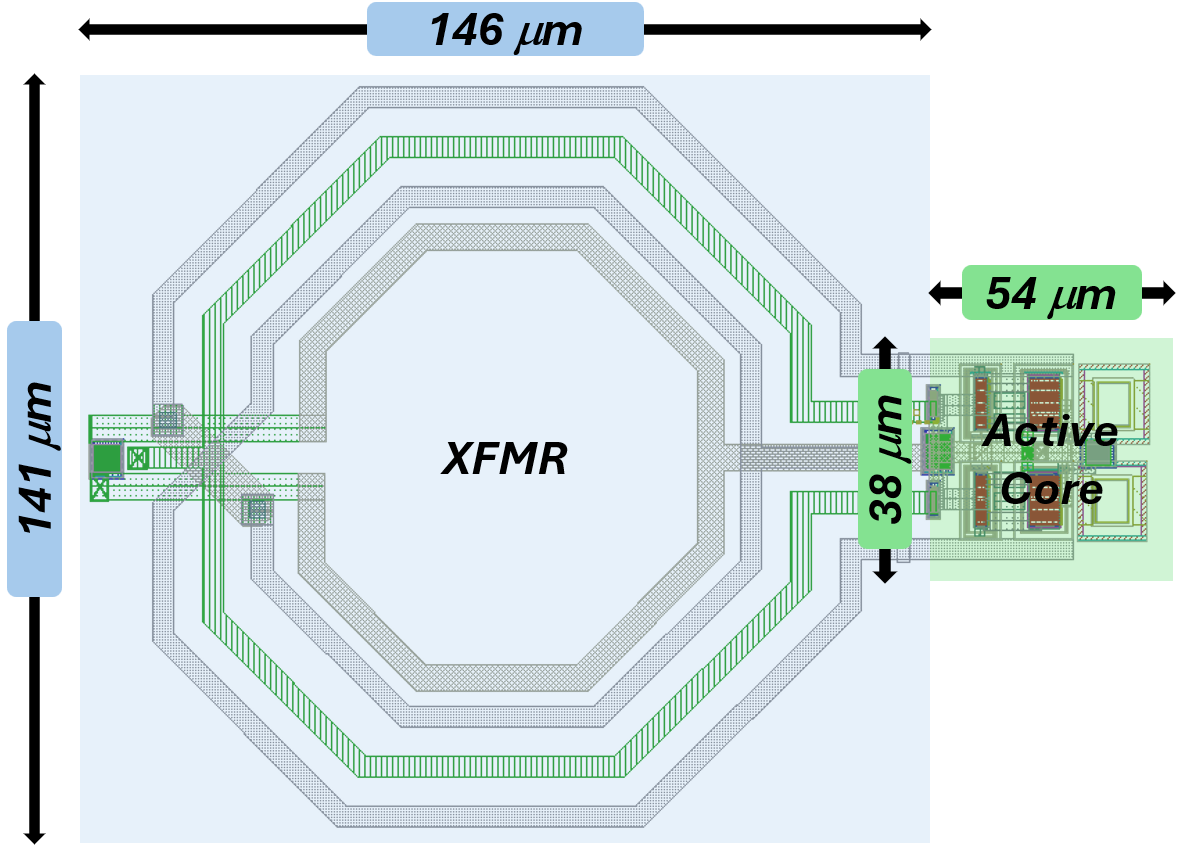}
    \caption{Layout of the proposed triple-coupled XFMR tank based VCO}
    \label{fig:Layout}
\end{figure}
\begin{figure}
    \centering
    \includegraphics[angle=0, width=\linewidth]{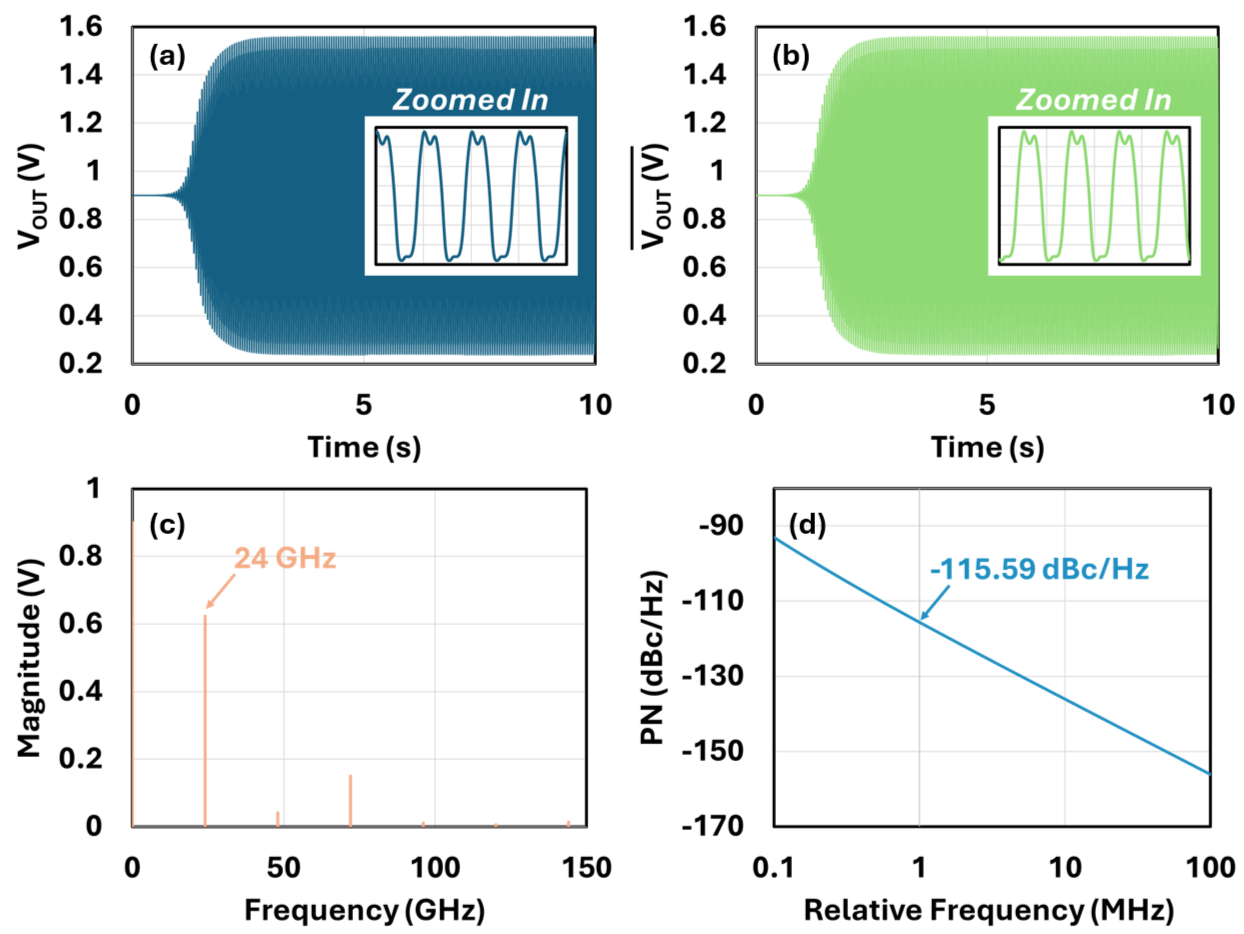}
    \caption{Transient waveforms of (a) $V_{out}$ and (b) $\overline{V_{out}}$, (c) Spectrum of $V_{out}$ and (d) Phase Noise profile}
    \label{fig:Transient_and_PSS_PN}
\end{figure}
\begin{figure*}
    \centering
    \includegraphics[width=\linewidth]{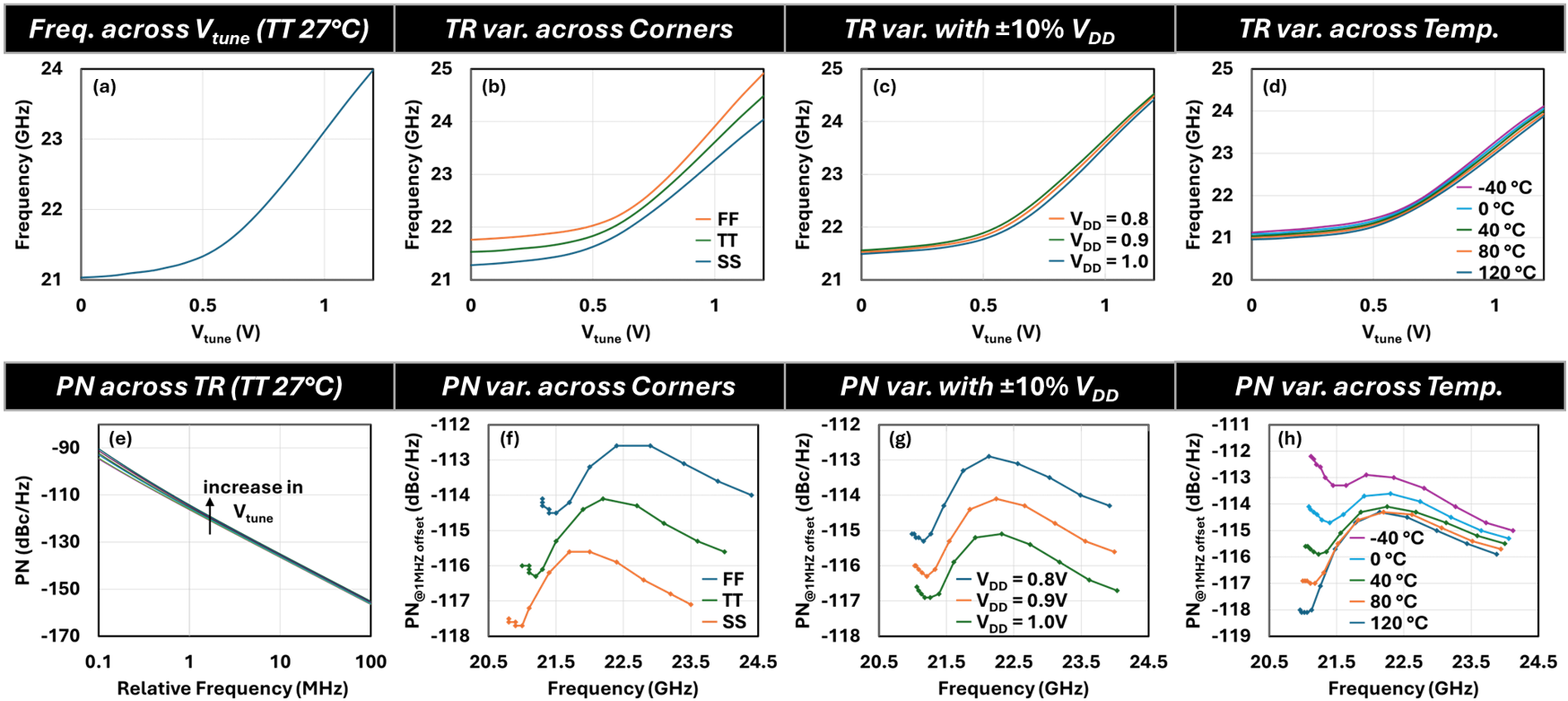}
    \caption{Post-layout simulation results summarizing TR and PN performance of the proposed VCO: (a) frequency vs $V_{\text{tune}}$ at TT 27\,$^\circ$C, (b)-(d) TR variation across PVT, respectively, (e) PN profile at TT 27\,$^\circ$C, and (f)-(h) PN variation across PVT, respectively.}
    \label{fig:Post_layout}
\end{figure*}
The proposed VCO is voltages biased, as it
provides more voltage headroom for the active-core MOSFETs
and eliminates a 
source of noise.
Varactors are incorporated to tune the 
fundamental frequency of the
VCO.
Various existing techniques 
tune both second and third harmonics to achieve low PN.
However, they typically require additional area.
As a trade-off, the second harmonic is not actively tuned in this work. 
The value of varactors are chosen such that it introduces a CM impedance along with $L_3$, even though it does not stay at $2f_0$ as $f_0$ gets tuned.
Instead, the NC active-core topology \cite{NC_8637955} is employed, which inherently provides wideband noise suppression.$N_{1,2}$ and $P_{1,2}$ on each side switch simultaneously within each half-cycle, enabling NC operation.

Fig. \ref{fig:Layout} shows the layout of the VCO.
The transistors $N_{1,2}$ are 
source degenerated by $P_{1,2}$,
with dimensions of $15~\mu\text{m}/0.6~\mu\text{m}$
and
$30~\mu\text{m}/0.6~\mu\text{m}$, respectively.
The VCO core, along with the associated capacitors and varactors, occupies an area of 
$0.02268~\text{mm}^2$.
LC 
Buffers are designed such that the output of the VCO output
remains stable across the 
TR.
Post-layout verification is carried out using
Cadence
SpectreRF, with electromagnetic effects incorporated through an 
S-parameter file 
generated using Cadence EMX. 
The VCO consumes 5.4 mA from a 0.9 V supply.
%
%
\begin{table*}[!t]
\centering
\caption{Performance summary and comparison with state-of-the-art mmWave VCOs}
\label{tab:comparison_table}
\resizebox{\textwidth}{!}{%
\begin{tabular}{|c|c|c|c|c|c|c|c|c|c|c|}
\hline 
\textbf{References} & \textbf{Tech. (nm)} & \textbf{Meas./Sim.} & \textbf{Frequency (GHz)} & \textbf{TR (\%)} & \textbf{Supply (V)} & \textbf{Power (mW)} & \textbf{PN (dBc/Hz)} & \textbf{Area (mm\textsuperscript{2})} & \textbf{FoM (dBc/Hz)} & \textbf{FoM\textsubscript{A} (dBc/Hz)}\\
\hline
\cite{two_port_2022_RuiMartin_mmWave_RLCM} & 65 & Meas. & 24.62 - 28.6 & 15.16 & 1 & 9.7 - 10.5 & -111.4 & 0.1924 & 189.4 & 196.56 \\
\hline
\cite{Reasonable_9085341} & 28 & Meas. & 19.5 & 12 & 0.9 & 20.7 & -112 & 0.07 & 185 & 196.56 \\
\hline
\cite{Self_Mixing_9738437} & 65 & Meas. & 20.7 - 28 & 29.98 & 1 & 12.65 - 15.1 & -107.9 & 0.1128 & 184.75 & 194.23 \\
\hline
\cite{NEWCAS_2023} & 65 & Sim. & 9.02 & 49.7 & 0.6 & 20 & -116.7* & 0.0492 & 184.1 - 185.9 & 195.88 \\
\hline
\cite{ISCAS_2023} & 40 & Sim. & 12.1 - 16.5 & 30.8 & 1.1 & 14.33 & -109 to -118 & 0.03 & 181.8 - 188.1 & 194.3 - 206.02 \\
\hline
\textbf{This Work} & \textbf{65} & \textbf{Sim.} & \textbf{23.99} & 13.15 & \textbf{0.9} & \textbf{5.4} & \textbf{-115.59} & \textbf{0.02268} & \textbf{196.86} & \textbf{212.31}  \\
\hline
\end{tabular}%
}
\begin{flushleft}
{\scriptsize *Interpolated from the graph}, 
{\scriptsize FoM $= -PN + 20\log_{10}\!(f_{o}/\Delta f) - 10\log_{10}\!(P_{DC}/1\,\mathrm{mW})$, $FoM_A=FoM+10\log_{10}(1\,\mathrm{mm}^2/Area)$}
\end{flushleft}
\vspace{-10pt}
\end{table*}
Fig. \ref{fig:Transient_and_PSS_PN} (a) and (b) show the transient waveforms of $V_{OUT}$ and $\overline{V_{OUT}}$. 
They clearly 
indicate the presence of higher-order
harmonic components in the oscillator output.
Fig. \ref{fig:Transient_and_PSS_PN} (c) shows the periodic steady state (PSS) spectrum of $V_{OUT}$ and (d) shows the corresponding PN curve.
We observe that the design achieves a PN of 
$-116.03$ dBc/Hz @ 1MHz offset 
at 24 GHz operating frequency.

Fig. \ref{fig:Post_layout} shows the detailed post-layout simulations of the proposed VCO.
Fig. \ref{fig:Post_layout}(a) shows frequency vs $V_{tune}$ (at TT, 27$^\circ$C) which depicts a TR of 3.96 GHz (21.03 - 23.99 GHz). 
Fig.  \ref{fig:Post_layout}(b) shows the frequency variation across 
SS, TT, and FF 
process corners. The maximum
frequency
difference between corners at the same $V_{tune}$
is 0.88 GHz. 
Fig. \ref{fig:Post_layout}(c)
and (d)
present the frequency variation with supply voltage (0.8–1.0~V) and temperature ($-40^\circ$C to $120^\circ$C), where the maximum differences observed across the full TR are 0.11 GHz and 0.24 GHz, respectively.
Overall, the oscillator maintains consistent tuning behavior under supply and temperature variations, 
while the frequency shift across process corners is comparatively larger.
Fig. \ref{fig:Post_layout}(e) shows the PN performance of the proposed design at TT 27$^\circ$C, where it is $<$-114.14 dBc/Hz @ 1 MHz offset, across the TR. It achieves a minimum PN of –116.25 dBc/Hz @ 1MHz offset, for a carrier frequency of 21.21 GHz.
Fig. \ref{fig:Post_layout}(f) shows the PN at 1 MHz offset across process corners, where the variation remains within 5 dB across the full TR. 
Fig. \ref{fig:Post_layout}(g) presents the PN variation with supply voltage (0.8-1.0~V), which stays within 4 dB. 
Fig. \ref{fig:Post_layout}(h) shows the temperature dependence from $-40\,^\circ$C to $120\,^\circ$C, where the variation is within approximately 5 dB. 

Table \ref{tab:comparison_table} summarizes the performance of the proposed VCO and compares it with recently reported mmWave VCO designs.
The post-layout results show that the proposed topology achieves PN performance comparable to prior works. At the same time, it exhibits the lowest power consumption and smallest area among the compared designs 
As a result, it achieves the highest FoM and FoM$_A$ in this comparison.

\section{Conclusion}

This work presents the analysis and design of a sixth-order 
triple-coupled XFMR tank enabling third-harmonic expansion for mmWave VCOs. 
A consistent closed-form derivation of the three DM resonance frequencies is developed, 
ensuring physically valid solutions across practical coupling conditions.
By aligning the higher-order modes around the third harmonic, wideband third-harmonic expansion is achieved without auxiliary low-Q harmonic-tuning elements. To avoid the area overhead of second-harmonic tuning structures, an NC active core is employed to limit the device noise coupled into the tank while maintaining PN performance. Implemented in TSMC 65-nm CMOS, the 21.03–23.99 GHz (13.5\%) TR, –115.58 dBc/Hz PN at 1 MHz
offset at 23.99 GHz, and peak FoM/FoMT/FoMA of 195.86/198.24/212.31 dBc/Hz. The VCO occupies 0.02268 mm$^2$ and consumes less than 5.4 mW from 0.9 V supply.

\section*{Acknowledgment}
The author would like to thank the Chips-to-Startup (C2S) program, Ministry of Electronics and Information Technology (MeitY), Govt. of India, I-Hub Data, IIIT Hyderabad, India, and IC-WiBES research group at IIIT Hyderabad for their support during this research.

\bibliographystyle{IEEEtran}
\bibliography{references}

@ARTICLE{Hajimiri_658619,
  author={Hajimiri, A. and Lee, T.H.},
  journal={IEEE Journal of Solid-State Circuits}, 
  title={{A general theory of phase noise in electrical oscillators}}, 
  year={1998},
  volume={33},
  number={2},
  pages={179-194},
  doi={10.1109/4.658619}}

@ARTICLE{Hegazi,
  author={Hegazi, E. and Sjoland, H. and Abidi, A.A.},
  journal={IEEE Journal of Solid-State Circuits}, 
  title={{A filtering technique to lower LC oscillator phase noise}}, 
  year={2001},
  volume={36},
  number={12},
  pages={1921-1930},
  doi={10.1109/4.972142}}

@ARTICLE{Murphy,
  author={Murphy, David and Darabi, Hooman and Wu, Hao},
  journal={IEEE Journal of Solid-State Circuits}, 
  title={{Implicit Common-Mode Resonance in LC Oscillators}}, 
  year={2017},
  volume={52},
  number={3},
  pages={812-821},
  doi={10.1109/JSSC.2016.2642207}}

@ARTICLE{two_port_2006,
  author={Huijung Kim and Seonghan Ryu and Yujin Chung and Jinsung Choi and Bumman Kim},
  journal={IEEE Transactions on Microwave Theory and Techniques}, 
  title={{A low phase-noise CMOS VCO with harmonic tuned LC tank}}, 
  year={2006},
  volume={54},
  number={7},
  pages={2917-2924},
  doi={10.1109/TMTT.2006.877439}}

@ARTICLE{two_port_2018,
  author={Guo, Hao and Chen, Yong and Mak, Pui-In and Martins, Rui P.},
  journal={IEEE Solid-State Circuits Letters}, 
  title={{A 0.083-mm2 25.2-to-29.5 GHz Multi-LC-Tank Class-F234 VCO With a 189.6-dBc/Hz FOM}}, 
  year={2018},
  volume={1},
  number={4},
  pages={86-89},
  doi={10.1109/LSSC.2018.2851499}}

@INPROCEEDINGS{two_port_2019,
  author={Guo, Hao and Chen, Yong and Mak, Pui-In and Martins, Rui P.},
  booktitle={2019 IEEE International Solid-State Circuits Conference - (ISSCC)}, 
  title={{26.2 A 0.08mm2 25.5-to-29.9GHz Multi-Resonant-RLCM-Tank VCO Using a Single-Turn Multi-Tap Inductor and CM-Only Capacitors Achieving 191.6dBc/Hz FoM and 130kHz 1/f3 PN Corner}}, 
  year={2019},
  volume={},
  number={},
  pages={410-412},
  doi={10.1109/ISSCC.2019.8662470}}

@INPROCEEDINGS{two_port_2020_RFIC,
  author={Guo, Hao and Chen, Yong and Mak, Pui-In and Martins, Rui P.},
  booktitle={2020 IEEE Radio Frequency Integrated Circuits Symposium (RFIC)}, 
  title={{A 0.082mm2 24.5-to-28.3GHz Multi-LC-Tank Fully-Differential VCO Using Two Separate Single-Turn Inductors and a 1D-Tuning Capacitor Achieving 189.4dBc/Hz FOM and 200±50kHz 1/f3 PN Corner}}, 
  year={2020},
  volume={},
  number={},
  pages={235-238},
  doi={10.1109/RFIC49505.2020.9218290}}

@ARTICLE{two_port_2020_TCAS1_RuiMartin_current_reuse,
  author={Huang, Yunbo and Chen, Yong and Guo, Hao and Mak, Pui-In and Martins, Rui P.},
  journal={IEEE Transactions on Circuits and Systems I: Regular Papers}, 
  title={{A 3.3-mW 25.2-to-29.4-GHz Current-Reuse VCO Using a Single-Turn Multi-Tap Inductor and Differential-Only Switched-Capacitor Arrays With a 187.6-dBc/Hz FOM}}, 
  year={2020},
  volume={67},
  number={11},
  pages={3704-3717},
  doi={10.1109/TCSI.2020.3013259}}

@ARTICLE{two_port_2022_RuiMartin_mmWave_RLCM,
  author={Guo, Hao and Chen, Yong and Yang, Chaowei and Mak, Pui-In and Martins, Rui P.},
  journal={IEEE Transactions on Circuits and Systems I: Regular Papers}, 
  title={{A Millimeter-Wave CMOS VCO Featuring a Mode-Ambiguity-Aware Multi-Resonant-RLCM Tank}}, 
  year={2022},
  volume={69},
  number={1},
  pages={172-185},
  doi={10.1109/TCSI.2021.3096196}}

@ARTICLE{Self_Mixing_9738437,
  author={Esmaeeli, Omid and Lightbody, Sam and Shirazi, Amir Hossein Masnadi and Djahanshahi, Hormoz and Zavari, Rod and Mirabbasi, Shahriar and Shekhar, Sudip},
  journal={IEEE Transactions on Circuits and Systems I: Regular Papers}, 
  title={{A Transformer-Based Technique to Improve Tuning Range and Phase Noise of a 20–28GHz LCVCO and a 51–62GHz Self-Mixing LCVCO}}, 
  year={2022},
  volume={69},
  number={6},
  pages={2351-2363},
  doi={10.1109/TCSI.2022.3158656}}

@ARTICLE{hr_classF,
  author={Babaie, Masoud and Staszewski, Robert Bogdan},
  journal={IEEE Journal of Solid-State Circuits}, 
  title={{A Class-F CMOS Oscillator}}, 
  year={2013},
  volume={48},
  number={12},
  pages={3120-3133},
  doi={10.1109/JSSC.2013.2273823}}

@ARTICLE{hr_classF-1,
  author={Lim, Chee Cheow and Ramiah, Harikrishnan and Yin, Jun and Mak, Pui-In and Martins, Rui P.},
  journal={IEEE Journal of Solid-State Circuits}, 
  title={{An Inverse-Class-F CMOS Oscillator With Intrinsic-High-Q First Harmonic and Second Harmonic Resonances}}, 
  year={2018},
  volume={53},
  number={12},
  pages={3528-3539},
  doi={10.1109/JSSC.2018.2875099}}

@ARTICLE{hr_1byf_upconversion,
  author={Shahmohammadi, Mina and Babaie, Masoud and Staszewski, Robert Bogdan},
  journal={IEEE Journal of Solid-State Circuits}, 
  title={{A 1/f Noise Upconversion Reduction Technique for Voltage-Biased RF CMOS Oscillators}}, 
  year={2016},
  volume={51},
  number={11},
  pages={2610-2624},
  doi={10.1109/JSSC.2016.2602214}}

@ARTICLE{hr_classF23_explicit_CM,
  author={Hu, Yizhe and Siriburanon, Teerachot and Staszewski, Robert Bogdan},
  journal={IEEE Journal of Solid-State Circuits}, 
  title={{A Low-Flicker-Noise 30-GHz Class-F23 Oscillator in 28-nm CMOS Using Implicit Resonance and Explicit Common-Mode Return Path}}, 
  year={2018},
  volume={53},
  number={7},
  pages={1977-1987},
  doi={10.1109/JSSC.2018.2818681}}

@ARTICLE{hr_classF23_optimal_Q,
  author={Hong, Feifan and Ding, Tianao and Zhao, Dixian},
  journal={IEEE Solid-State Circuits Letters}, 
  title={{A 196.5 dBc/Hz FOMT 16.8–21.6-GHz Class-F23 CMOS VCO With Transformer-Based Optimal Q-Factor Tank}}, 
  year={2022},
  volume={5},
  number={},
  pages={62-65},
  doi={10.1109/LSSC.2022.3160488}}

@ARTICLE{hr_VCO_Tutorial,
  author={Chen, Yong and Mak, Pui-In and Martins, Rui P.},
  journal={IEEE Transactions on Circuits and Systems II: Express Briefs}, 
  title={{High-Performance Harmonic-Rich Single-Core VCO With Multi-LC Tank: A Tutorial}}, 
  year={2022},
  volume={69},
  number={7},
  pages={3115-3121},
  doi={10.1109/TCSII.2022.3180351}}

@ARTICLE{hr_inverse_classF23,
  author={Du, Jianglin and Hu, Yizhe and Siriburanon, Teerachot and Kobal, Enis and Quinlan, Philip and Zhu, Anding and Staszewski, Robert Bogdan},
  journal={IEEE Journal of Solid-State Circuits}, 
  title={{A Compact 0.2–0.3-V Inverse-Class-F23 Oscillator for Low 1/f3 Noise Over Wide Tuning Range}}, 
  year={2022},
  volume={57},
  number={2},
  pages={452-464},
  doi={10.1109/JSSC.2021.3098770}}

@INPROCEEDINGS{hr_trifilar,
  author={Cao, Hanzhang and Huang, Tongde and Liu, Xiaolong and Wang, Hao and Jin, Jin and Wu, Wen},
  booktitle={2023 IEEE Asian Solid-State Circuits Conference (A-SSCC)}, 
  title={{A 5.2GHz Trifilar Transformer-Based Class-F23 Noise Circulating VCO with FoM of 192.6 dBc/Hz}}, 
  year={2023},
  volume={},
  number={},
  pages={1-3},
  doi={10.1109/A-SSCC58667.2023.10347969}}

@ARTICLE{Reasonable_9085341,
  author={Franceschin, Alessandro and Andreani, Pietro and Padovan, Fabio and Bassi, Matteo and Bevilacqua, Andrea},
  journal={IEEE Journal of Solid-State Circuits}, 
  title={{A 19.5-GHz 28-nm Class-C CMOS VCO, With a Reasonably Rigorous Result on 1/f Noise Upconversion Caused by Short-Channel Effects}}, 
  year={2020},
  volume={55},
  number={7},
  pages={1842-1853},
  doi={10.1109/JSSC.2020.2987702}}

@ARTICLE{NC_8637955,
  author={Wang, Fei and Wang, Hua},
  journal={IEEE Journal of Solid-State Circuits}, 
  title={{A Noise Circulating Oscillator}}, 
  year={2019},
  volume={54},
  number={3},
  pages={696-708},
  doi={10.1109/JSSC.2018.2886321}}

@ARTICLE{hr_classF-1_2023,
  author={Meng, Xi and Li, Haoran and Chen, Peng and Yin, Jun and Mak, Pui-In and Martins, Rui P.},
  journal={IEEE Transactions on Circuits and Systems I: Regular Papers}, 
  title={{Analysis and Design of a 15.2-to-18.2-GHz Inverse-Class-F VCO With a Balanced Dual-Core Topology Suppressing the Flicker Noise Upconversion}}, 
  year={2023},
  volume={70},
  number={12},
  pages={5110-5123},
  doi={10.1109/TCSI.2023.3312817}}

@INPROCEEDINGS{NEWCAS_2023,
  author={Sachdeva, Ritesh and Kumar, Abhishek},
  booktitle={2023 21st IEEE Interregional NEWCAS Conference (NEWCAS)}, 
  title={{A Single-Turn Inductor based Compact and Wide-Tuning LC-VCO using Dual-Resonant Modes}}, 
  year={2023},
  volume={},
  number={},
  pages={1-5},
  doi={10.1109/NEWCAS57931.2023.10198205}}

@INPROCEEDINGS{ISCAS_2023,
  author={Jing, Jiayu and Li, Wei and Yuan, Ren and Xu, Hongtao},
  booktitle={2023 IEEE International Symposium on Circuits and Systems (ISCAS)}, 
  title={{A 12.1-16.5GHz Resistance Self-biased Inverse Class-F23 VCO Achieving 20-54kHz 1/f3 Corner Frequency}}, 
  year={2023},
  volume={},
  number={},
  pages={1-5},
  doi={10.1109/ISCAS46773.2023.10181361}}

@ARTICLE{classF23_6th_order,
  author={Tian, Shuo and Liu, Xiaolong},
  journal={IEEE Transactions on Circuits and Systems I: Regular Papers}, 
  title={{A Class-F23 CMOS Oscillator With Second and Third Harmonic Resonances Expansion}}, 
  year={2026},
  volume={73},
  number={1},
  pages={87-99},
  doi={10.1109/TCSI.2025.3583754}}

@ARTICLE{lowQ_9794818,
  author={Hong, Feifan and Zhang, Hao and Zhao, Dixian},
  journal={IEEE Transactions on Circuits and Systems I: Regular Papers}, 
  title={{An X-Band CMOS VCO Using Ultra-Wideband Dual Common-Mode Resonance Technique}}, 
  year={2022},
  volume={69},
  number={9},
  pages={3579-3590},
  doi={10.1109/TCSI.2022.3181139}}

@ARTICLE{lowQ_10130821,
  author={Lin, Ziyi and Jia, Haikun and Ma, Ruichang and Deng, Wei and Wang, Zhihua and Chi, Baoyong},
  journal={IEEE Journal of Solid-State Circuits}, 
  title={{A Low-Phase-Noise VCO With Common-Mode Resonance Expansion and Intrinsic Differential 2nd-Harmonic Output Based on a Single Three-Coil Transformer}}, 
  year={2024},
  volume={59},
  number={1},
  pages={253-267},
  doi={10.1109/JSSC.2023.3274178}}

@INPROCEEDINGS{lowQ_9365761,
  author={Guo, Hao and Chen, Yong and Mak, Pui-In and Martins, Rui P.},
  booktitle={2021 IEEE International Solid-State Circuits Conference (ISSCC)}, 
  title={{20.1 A 5.0-to-6.36GHz Wideband-Harmonic-Shaping VCO Achieving 196.9dBc/Hz Peak FoM and 90-to-180kHz 1/f3 PN Corner Without Harmonic Tuning}}, 
  year={2021},
  volume={64},
  number={},
  pages={294-296},
  doi={10.1109/ISSCC42613.2021.9365761}}

@ARTICLE{6th_order_tank_OG_10068119,
  author={Wang, Zhipeng and Ma, Kaixue and Ma, Zonglin and Shi, Hao and Fu, Haipeng and Xu, Jiangtao},
  journal={IEEE Transactions on Microwave Theory and Techniques}, 
  title={{A Reconfigurable Injection-Locked LO Generator With a Wideband-Harmonic-Shaping Class-F23 VCO for Multibands 5G mm-Wave}}, 
  year={2023},
  volume={71},
  number={9},
  pages={4144-4157},
  doi={10.1109/TMTT.2023.3251102}}

@INPROCEEDINGS{head_resonator_9365761,
  author={Guo, Hao and Chen, Yong and Mak, Pui-In and Martins, Rui P.},
  booktitle={2021 IEEE International Solid-State Circuits Conference (ISSCC)}, 
  title={{20.1 A 5.0-to-6.36GHz Wideband-Harmonic-Shaping VCO Achieving 196.9dBc/Hz Peak FoM and 90-to-180kHz 1/f3 PN Corner Without Harmonic Tuning}}, 
  year={2021},
  volume={64},
  number={},
  pages={294-296},
  doi={10.1109/ISSCC42613.2021.9365761}}

@ARTICLE{Dual_CM_9794818,
  author={Hong, Feifan and Zhang, Hao and Zhao, Dixian},
  journal={IEEE Transactions on Circuits and Systems I: Regular Papers}, 
  title={{An X-Band CMOS VCO Using Ultra-Wideband Dual Common-Mode Resonance Technique}}, 
  year={2022},
  volume={69},
  number={9},
  pages={3579-3590},
  doi={10.1109/TCSI.2022.3181139}}

@ARTICLE{aux_coil_10130821,
  author={Lin, Ziyi and Jia, Haikun and Ma, Ruichang and Deng, Wei and Wang, Zhihua and Chi, Baoyong},
  journal={IEEE Journal of Solid-State Circuits}, 
  title={{A Low-Phase-Noise VCO With Common-Mode Resonance Expansion and Intrinsic Differential 2nd-Harmonic Output Based on a Single Three-Coil Transformer}}, 
  year={2024},
  volume={59},
  number={1},
  pages={253-267},
  doi={10.1109/JSSC.2023.3274178}}

@ARTICLE{switched_XFMR_10274730,
  author={Li, Yubing and Huang, Zemeng and Song, Linying and Yang, Taojun and Li, Xiuping},
  journal={IEEE Transactions on Microwave Theory and Techniques}, 
  title={{An X-Band Low-Phase-Noise Class-F23 VCO Without Manual Harmonic Tuning Based on Switched-Transformer and Wideband Common-Mode Resonance}}, 
  year={2024},
  volume={72},
  number={5},
  pages={3076-3090},
  doi={10.1109/TMTT.2023.3319988}}

@ARTICLE{multi_LC_10068119,
  author={Wang, Zhipeng and Ma, Kaixue and Ma, Zonglin and Shi, Hao and Fu, Haipeng and Xu, Jiangtao},
  journal={IEEE Transactions on Microwave Theory and Techniques}, 
  title={{A Reconfigurable Injection-Locked LO Generator With a Wideband-Harmonic-Shaping Class-F23 VCO for Multibands 5G mm-Wave}}, 
  year={2023},
  volume={71},
  number={9},
  pages={4144-4157},
  doi={10.1109/TMTT.2023.3251102}}

\end{document}